\documentclass[12pt]{iopart}
\usepackage{graphicx}
\begin{document}
\title {Quantum Brownian motion under rapid periodic forcing}
\author{ Malay Bandyopadhyay and Mustansir Barma}
\vskip 0.5cm
\address{Department of Theoretical Physics, Tata Institute of Fundamental Research, Homi Bhabha Road, Colaba, Mumbai-400005, India.}
\vskip 0.5cm
\begin{abstract}
We study the steady state behaviour of a confined quantum Brownian particle subjected to a space-dependent, rapidly oscillating time-periodic force. To leading order in the period of driving, the result of the oscillating force is an effective static potential which has a quantum dissipative contribution, $V_{QD}$, which adds on to the classical result. This is shown using a coherent state representation of bath oscillators. $V_{QD}$ is evaluated exactly in the case of an Ohmic dissipation bath. It is strongest for intermediate values of the damping, where it can have pronounced effects.
\end{abstract}
\maketitle
{\section {Introduction}}
Parametric phenomena, induced by giving a time-dependence to a parameter or coupling constant, are ubiquitous in physics.  Applications range from confining ions in quadrupolar traps (Paul traps) \cite{paul} to controlling particle bunching and dilution in particle accelerators \cite{balandin}.\\ 
\indent
When the parameter in question oscillates rapidly, certain simplifications occur, as first demonstrated by Kapitza for the case of a simple pendulum whose point of oscillation is vibrated rapidly \cite{kapitza}. Kapitza's treatment for the pendulum was generalized by Landau and Lifshitz to apply to any forced system evolving through Newtonian dynamics, provided the forcing depends on the spatial coordinate and oscillates rapidly enough in time \cite{landau,perceival}. The result is simple and elegant. To leading order in the period of driving, the system feels an additional effective static potential; the form of this Kapitza-Landau-Lifshitz ($KLL$) potential, $V_{KLL}$, depends on the spatial variation of the forcing.\\
\indent 
Recently, this treatment has been generalized to the case of a Brownian particle subjected to similar forcing \cite{barma,malay}. The result of classical dissipation ($CD$), to leading order, is again an effective static potential, but with additional Wronskian terms, $V_{CD}$, reflecting the effect of dissipation in this system. To next order, it is shown that the result cannot be written as an effective potential, except in the limit of high damping \cite{barma}. An approximate Langevin approach used in \cite{malay}, leads to a result which agrees with the correct one to leading order, and to the next order if the damping is very large.\\
\indent 
The purpose of the present paper is to address this problem for a {\em  quantum dissipative system}. In the absence of dissipation, Cook {\textit et al.} \cite{cook} showed that to leading order a Schr\"{o}dinger equation with a rapid periodic driving term is tantamount to one with an additional static potential of the $KLL$ form, while Rahav {\textit et al} used an expansion in powers of the period to show that to the next order, the effective static Hamiltonian involves the particle coordinate and momentum in a non-separable way \cite{rahav}. In this paper, we model quantum dissipative ($QD$) Brownian motion by considering a particle coupled to an infinite set of harmonic oscillators \cite{zwangig,kac,legg1,weiss}. We are interested in steady state properties, which are independent of the initial condition. Using an initial coherent state of bath oscillators, a $c$-number generalized quantum Langevin equation (GQLE) is derived. We show that to leading order, the rapid forcing is tantamount to an effective static potential which can be written as $V_{KLL}+V_{CD}+V_{QD}$. Here $V_{QD}$ is an explicitly quantum dissipative contribution to the potential, which vanishes in the classical limit \cite{barma}. In the complete absense of dissipation, the result reduces to $V_{KLL}$ in agreement with \cite{cook,rahav}. $V_{QD}$ has most pronounced effects at intermediate values of the damping. In that regime, $V_{QD}$ is found to strongly influence the shape of the potential, e.g. the number and nature of the extrema of the potential.  \\
\section{Model and Formalism}
We consider a particle coupled to a model heat bath of the Zwanzig form \cite{zwangig}, described by the following Hamiltonian :
\begin{eqnarray}
\hat{\cal{H}} = \frac{\hat{p}^2}{2m}+V_0(\hat{x})+V_{ext}(\hat{x},t)+\sum_j\Big\lbrack\frac{\hat{p}_j^2}{2m_j}+\frac{1}{2}m_j\omega_j^2\Big(\hat{q}_j-\frac{c_j}{m_j\omega_j^2}\hat{x}\Big)^2\big\rbrack, 
\end{eqnarray}
where $\hat{x}$ and $\hat{p}$ are coordinate and momentum operators of the Brownian particle, $\lbrace \hat{q}_j, \hat{p}_j\rbrace$ is the set of co-ordinate and momentum operators for the heat-bath oscillators, and $V_0(\hat{x})$ is the harmonic confining potential. The potential $V_{ext}(\hat{x},t)$ arises from the external force field. The co-ordinate and momentum operators follow the usual commutation relations
\begin{eqnarray}
\lbrack\hat{x},\hat{p}\rbrack = i\hbar, \ \ \ \ \ \
\lbrack \hat{q}_j,\hat{p}_k\rbrack =i\hbar \delta_{jk}.
\end{eqnarray}
\indent
Eliminating the reservoir degrees of freedom \cite{kac,weiss}, we obtain the operator GQLE corresponding to Hamiltonian (1) :
\begin{eqnarray}
m\ddot{\hat{x}}(t)+m\int_0^tdt_1\gamma(t-t_1)\dot{\hat{x}}(t_1)+V_0^{\prime}(\hat{x})
= F_{ext}(\hat{x},t)+\hat{\eta}(t),
\end{eqnarray}
where noise operator $\hat{\eta}(t)$, memory kernel $\gamma(t)$ and external force $F_{ext}(\hat{x},t)$ are given by 
\begin{eqnarray}
\hat{\eta}(t)= \sum_j c_j\Big\lbrack\Big(\hat{q}_j(0)-\frac{c_j}{m_j\omega_j^2}\hat{x}(0)\Big)\cos(\omega_jt)+\frac{\hat{p}_j(0)}{m_j\omega_j}\sin(\omega_jt)\Big\rbrack,\\
\gamma(t)=\sum_j\frac{c_j^2}{m_j\omega_j^2}\cos(\omega_j t),\\
F_{ext}(\hat{x},t)=-V^{\prime}_{ext}(\hat{x},t).
\end{eqnarray}
\indent
Our aim is to replace the exact GQLE in operator form  Eq. (3) by its $c$-number equivalent. To achieve this, we follow the method of Banerjee {\textit et al.} \cite{banerjee1,banerjee2} and introduce product separable quantum states to describe the initial $t=0$ state of the particle and the bath oscillators,
\begin{equation} 
|\psi_{initial}\rangle=|\phi\rangle\lbrace|\alpha_j\rangle\rbrace,
\end{equation}
where $|\phi\rangle$ denotes any arbitrary initial state of the particle and $|\alpha_j\rangle$ corresponds to the initial coherent state of the $j$th oscillator. Here $|\alpha_j\rangle$ is given by
\begin{equation}
|\alpha_j\rangle = \exp(-\frac{|\alpha_j^2|}{2})\sum_{n_j=0}^{\infty}(\frac{\alpha_j^{n_j}}{\sqrt{n_j!}})|n_j\rangle,
\end{equation}
and $\alpha_j$ is expressed in terms of the variables of the coordinate and momentum of the $j$th  oscillator 
\begin{eqnarray}
\langle\hat{q}_j(0)\rangle-\langle\hat{x}(0)\rangle  = \sqrt{\frac{\hbar\omega_j^{-1}}{2}} \Big(\alpha_j^*+\alpha_j\Big)\\
\langle\hat{p}_j(0)\rangle  = i\sqrt{\frac{\hbar\omega_j}{2}}\Big(\alpha_j^*-\alpha_j\Big).
\end{eqnarray}
Following \cite{banerjee1,banerjee2} and doing the quantum-statistical averaging starting from the intial product separable quantum state (Eq. 7), one obtains the $c$-number GQLE :
\begin{eqnarray}
\hskip-1.8cm
m\ddot{x}(t)+m\int_0^tdt_1\gamma(t-t_1)\dot{x}(t_1)+V_0^{\prime}(x)+V_{ext}^{\prime}(x,t)&=&\eta(t)+Q_0(x,t)\nonumber\\
&+&Q_{ext}(x,t),
\end{eqnarray}
where 
\begin{eqnarray}
x(t)=\langle\hat{x}\rangle, \nonumber \\
\hskip-1.8cm
\eta(t)=\langle\hat{\eta}(t)\rangle=\sum_j\lbrack\lbrace\langle\hat{q}_j(0)\rangle-\langle\hat{x}(0)\rangle\rbrace \frac{c_j}{m_j\omega_j^2}\cos(\omega_jt)+\sqrt{\frac{c_j}{m_j\omega_j^2}}\langle\hat{p}_j(0)\rangle\sin(\omega_jt),
\end{eqnarray}
and $Q_0$ and $Q_{ext}$ are quantum fluctuation terms, given by :
\begin{eqnarray}
Q_0(x,t)=V_0^{\prime}(x)-\langle V_0^{\prime}(\hat{x})\rangle, \\
Q_{ext}(x,t)=V_{ext}^{\prime}(x,t)-\langle V_{ext}^{\prime}(\hat{x},t)\rangle.
\end{eqnarray}
\indent
In order that $\eta(t)$ be an effective $c$-number noise, we must have
\begin{eqnarray}
\langle\eta(t)\rangle_s=0, \\
\langle\eta(t)\eta(t^{\prime})\rangle_s=\frac{1}{2}\sum_j\frac{c_j}{m_j\omega_j^2}\hbar\omega_j\coth(\frac{\hbar\omega_j}{2k_BT})\cos\omega_j(t-t^{\prime}),
\end{eqnarray}
where $\langle .... \rangle_s$ denotes statistical average over the initial distribution of the mean values of the momenta and co-ordinates of the bath oscillators. Equations (15) and (16) imply that $\xi(t)$ is centered around zero and satisfies the quantum fluctuation-dissipation relation, and are obtained if and only if the initial quantum mechanical mean values of momenta and co-ordinates of the bath oscillators have the following distribution \cite{banerjee1,banerjee2}:
\begin{eqnarray}
P_j = \exp\Big\lbrack-\frac{\omega_j^2\lbrace\langle\hat{q}_j(0)\rangle-\langle\hat{x}(0)\rangle\rbrace^2+\langle\hat{p}_j(0)\rangle^2}{2\hbar\omega_j(\bar{n}_j+\frac{1}{2})}\Big\rbrack,
\end{eqnarray}
where $\bar{n}_j=\lbrack\exp(\frac{\hbar\omega_j}{k_BT})-1\rbrack^{-1}$, the average thermal phonon number at temperature $T$. Thus, the statistical average of any quantum mechanical mean value $O_j\Big(\langle\hat{p}_j(0)\rangle,\lbrace\langle\hat{q}_j(0)\rangle-\langle\hat{x}(0)\rangle\rbrace\Big)$ is defined as
\begin{eqnarray}
\hskip-1.8cm
\langle O_j\rangle_s&=&\int O_j\Big(\langle\hat{p}_j(0)\rangle,\lbrace\langle\hat{q}_j(0)\rangle-\langle\hat{x}(0)\rangle\rbrace\Big) P_j d\langle\hat{p}_j(0)\rangle d\lbrace\langle\hat{q}_j(0)\rangle-\langle\hat{x}(0)\rangle\rbrace.
\end{eqnarray}
Using equations (12), (17), and (18), one can easily show the properties (15) and (16) of the c-number noise. $P_j$ is a canonical Wigner distribution for a displaced harmonic oscillator and always remains  positive \cite{wigner}. Now, one can interpret equation (11) as a $c$-number GQLE which is governed by a $c$-number noise $\eta(t)$ originating from the heat bath characterized by the properties (15) and (16). The two quantum fluctuation terms $Q_0(x,t)$ and $Q_{ext}(x,t)$ originate from the nonlinearity of the confining potential $V_0(x)$ and the externally applied rapidly oscillating potential $V_{ext}(x,t)$, respectively.\\
\section{Quantum fluctuation terms}
In this section, we discuss the derivation of the quantum terms $Q_0(X,t)$ and $Q_{ext}(X,t)$. Based on the quantum nature of the system, one can write
\begin{eqnarray}
&&\delta\hat{x}(t)=\hat{x}(t)-x(t), \\
&&\delta\hat{p}(t)=\hat{p}(t)-p(t),
\end{eqnarray}
where $x(t)=\langle\hat{x}(t)\rangle$, $p(t)= \langle\hat{p}(t)\rangle$.
Then, making a Taylor series expansion around $x$ in Eqs. (13) and (14), one obtains
\begin{eqnarray}
Q_0(x,t)&=&-\sum_{n\geq 2}\frac{1}{n!}V_0^{(n+1)}(x)\langle\delta\hat{x}^n\rangle,\\
Q_{ext}(x,t)&=&-\sum_{n\geq 2}\frac{1}{n!}V_{ext}^{(n+1)}(x,t)\langle\delta\hat{x}^n\rangle,
\end{eqnarray}
where $V_0^{(n+1)}(x)$ and $V_{ext}^{(n+1)}(x,t)$ are the $(n+1)$th derivatives of $V_0(x)$ and $V_{ext}(x,t)$. The terms $Q_0(x,t)$ and $Q_{ext}(x,t)$ depend on the quantum factors $\langle\delta\hat{x}^n(t)\rangle$ whose evaluation is discussed below.\\
\indent
Substituting Eqs. (19) and  (20) into Eq. (3), we obtain
\begin{eqnarray}
\hskip-1.8cm
m\delta\ddot{\hat{x}}+m\int_0^tdt_1\gamma(t-t_1)\delta\dot{\hat{x}}+V_0^{\prime\prime}(x)\delta\hat{x}+V_{ext}^{\prime\prime}(x,t)\delta\hat{x}
+\sum_{n\geq 2}\frac{1}{n!}V_0^{(n+1)}(x)\langle\delta\hat{x}^n\rangle \nonumber \\
+\sum_{n\geq 2}\frac{1}{n!}V_{ext}^{(n+1)}(x,t)\langle\delta\hat{x}^n\rangle=\delta\hat{\eta}(t),
\end{eqnarray}
where $\delta\hat{\eta}(t)=\hat{\eta}(t)-\eta(t)$.\\
\indent
We now consider the case of a harmonic confining potential $V_0(x)=\frac{1}{2}m\omega_0^2x^2$, in which case the derivatives in Eq. (21) vanish, implying $Q_0(x,t)=0$. With rapid driving, excursions around the smooth part are small, so that for our purpose of deriving the leading order term, we may replace $\langle\delta\hat{x}^2(t)\rangle$ by $\langle\delta\hat{x}^2(t)\rangle_0$ where $\langle ....\rangle_0$ denotes an average in the absence of driving. Thus, we have 
\begin{eqnarray}
Q_{ext}(x,t)\simeq-\frac{1}{2!}V_{ext}^{\prime\prime\prime}(x,t)\langle\delta\hat{x}^2(t)\rangle_0.
\end{eqnarray}
Equation (23) then reduces to 
\begin{equation}
m\delta\ddot{\hat{x}}+m\int_0^tdt_1\gamma(t-t_1)\delta\dot{\hat{x}}+m\omega_0^2\delta\hat{x}=\delta\hat{\xi}.
\end{equation}
Equation (25) may be solved by Laplace transformations, leading to
\begin{equation}
\delta\hat{x}(t)=G(t)\delta\hat{x}(0)+H(t)\delta\dot{\hat{x}}(0)+\int_0^tdt^{\prime}H(t-t^{\prime})\delta\hat{\xi}(t^{\prime}),
\end{equation}
where $G(t)$ and $H(t)$ are the inverse Laplace transforms of $\tilde{G}(s)$ and $\tilde{H}(s)$ respectively given by
\begin{eqnarray}
\tilde{G}(s)=\frac{s+\tilde{\gamma}(s)}{s^2+s\tilde{\gamma}(s)+\omega_0^2},\\
\tilde{H}(s)=\frac{1}{s^2+s\tilde{\gamma}(s)+\omega_0^2}.
\end{eqnarray}
Here
\begin{equation}
\tilde{\gamma}(s)=\int_0^{\infty}\gamma(t)e^{-st}dt
\end{equation}
is the Laplace transform of the Frictional kernel $\gamma(t)$. Equation (26) then leads to
\begin{eqnarray}
\hskip-1.8cm
\langle\delta\hat{x}^2(t)\rangle_0&=&{G}^2(t)\langle\delta\hat{x}^2(0)\rangle_0+\frac{H^2(t)}{m^2}\langle\delta\hat{p}^2(0)\rangle_0+\frac{{G}(t)H(t)}{m}\langle\delta\hat{x}(0)\delta\hat{p}(0)+\delta\hat{p}(0)\delta\hat{x}(0)\rangle_0 \nonumber \\
\hskip-1.8cm
&&+2\int_0^t dt^{\prime}\int_0^{t^{\prime}}dt^{\prime\prime}H(t-t^{\prime})H(t-t^{\prime\prime})\langle\delta\hat{\eta}(t^{\prime})\delta\hat{\eta}(t^{\prime\prime})\rangle.
\end{eqnarray}
\indent  
Let us choose initial conditions corresponding to minimum uncertainty states \cite{banerjee1,sundaram}:, so that $\langle\delta\hat{x}^2(0)\rangle_0=\frac{\hbar}{2m\omega_0}$, $\langle\delta\hat{p}^2(0)\rangle_0=\frac{m\hbar\omega_0}{2}$, and $\langle\delta\hat{x}(0)\delta\hat{p}(0)+\delta\hat{p}(0)\delta\hat{x}(0)\rangle_0=\hbar$.
In order to find $H(t)$ and $G(t)$ (or equivalently $\tilde{H}(s)$ and $\tilde{G}(s)$), we need to know $\gamma(t)$. We make the customary choice of an Ohmic heat bath, which leads to $\gamma(t)=\gamma_0\delta(t)$ and finally to an explicit forms for $H(t)$ and $G(t)$. In the underdamped regime ($\omega_0 > \gamma /2$), we find 
\begin{eqnarray}
H(t)=\frac{1}{\omega_1}e^{-\frac{\gamma_0}{2}t}\sin(\omega_1 t),\\
G(t)=e^{-\frac{\gamma_0}{2}t}\Big\lbrack\cos(\omega_1t)+\frac{\gamma_0}{2m\omega_1}\sin(\omega_1 t)\Big\rbrack.
\end{eqnarray}
where $\omega_1=\pm\sqrt{\omega_0^2-\frac{\gamma_0^2}{4}}$. For the overdamped case ($\omega_0 < \frac{\gamma}{2}$), $\omega_1$ becomes imaginary and $H(t)$ and $G(t)$ are modified to
\begin{eqnarray}
H(t)=\frac{1}{\omega_1^{\prime}}e^{-\frac{\gamma_0}{2}t}\sinh(\omega_1^{\prime}t),\\
G(t)=e^{-\frac{\gamma_0}{2}t}\Big\lbrack\cosh(\omega_1^{\prime}t)+\frac{\gamma_0}{2m\omega_1^{\prime}}\sinh(\omega_1^{\prime}t)\Big\rbrack,
\end{eqnarray}
where $\omega_1^{\prime}=\pm\sqrt{\frac{\gamma_0^2}{4}-\omega_0^2}$.\\
\indent
With the Ohmic condition for the heat bath, $\gamma(t)=\gamma_0\delta(t)$, the double integral in equation (30) can be evaluated with the result 
$$\frac{\gamma_0}{m\pi}\int_0^{\infty}d\omega \hbar\omega\coth\Big(\frac{\hbar\omega}{2k_BT}\Big)\left|\frac{1-e^{-(\frac{\gamma_0}{2}-i\omega)t}\lbrack\cos\omega_1 t+ (\frac{\gamma_0}{2}-i\omega_1)\frac{\sin\omega_1 t}{\omega_1}\rbrack}{\omega^2-\omega_0^2+i\gamma_0\omega}\right|^2.$$ 
It reduces to a simple form for time large compared to $\gamma_0^{-1}$ : 
\begin{equation}
\langle \delta\hat{x}^2\rangle_{0}=\frac{\gamma_0}{m\pi}\int_0^{\infty}d\omega\hbar\omega\coth\Big(\frac{\hbar\omega}{2k_BT}\Big)\frac{1}{(\omega^2-\omega_0^2)^2+\gamma_0^2\omega^2}.
\end{equation}
In the limit of weak damping ($\gamma_0 << \omega_0$), Eq. (35) reduces to
\begin{equation}
\langle\delta\hat{x}^2\rangle_{0}=\frac{\hbar}{2m\omega_0}\coth\Big(\frac{\hbar\omega_0}{2\pi k_BT}\Big).
\end{equation}
while with strong damping ($\gamma_0 >> \omega_0$), we obtain 
\begin{eqnarray}
\hskip-1.8cm
\langle\delta\hat{x}^2\rangle_{0}=\frac{\hbar}{\pi\gamma_0}\Big\lbrack 2\ln(\frac{\gamma_0}{m\omega_0})+\ln(\frac{m\hbar\omega_0^2}{2\pi\gamma_0 k_BT})
+\frac{\pi\gamma_0 k_BT}{m\hbar\omega_0^2}-\Big(\frac{\pi m k_BT}{\hbar\gamma_0}\Big)^2+\gamma_E\Big\rbrack,
\end{eqnarray}
where $\gamma_E \simeq 0.57772$ is the Euler's constant. When substituted in Eq. (24), these answers for $\langle\delta\hat{x}^2\rangle_{0}$ determine the leading order form of the quantum fluctuation term Eq. (22).  
\section{The effective potential}
In this section we show that on time scales larger than the period of driving, the leading order efffect is to produce an effective potential which has an additive quantum dissipative term $V_{QD}$ whose form is derived below (Eq. 50).\\
\indent
For the harmonic confining potential and Ohmic dissipative bath, we have seen that the operator GQLE reduces to 
\begin{equation}
m\ddot{x}(t)+m\gamma_0\dot{x}(t)+V_0^{\prime}(x)=F_{ext}(x,t)+Q_{ext}(x,t)+\eta(t).
\end{equation}
If $\eta(t)$ is delta correlated as it is for an Ohmic dissipative bath, one can show that the noise term does not enter the effective Hamiltonian to leading order \cite{dattagupta}. Now, following the KLL approach \cite{landau} one can write
\begin{equation}
x(t)=X(t)+\xi(X,t),
\end{equation}
where $X(t)$ and $\xi (X,t)$ are the slow coordinate and rapidly oscillating coordinate respectively.  Inserting Eq. (39) in Eq. (38), we obtain 
\begin{eqnarray}
\hskip-0.8cm
m\ddot{X}(t)+m\ddot{\xi}&=&-\gamma_0\dot{X}-\gamma_0\dot{\xi}-\frac{\partial V_0(X)}{\partial X}-\xi\frac{\partial^2V_0}{\partial X^2}\nonumber \\
\hskip-0.8cm
&&+F_{ext}(X,t)+\xi F_{ext}^{\prime}(X,t)+Q_{ext}(X,t)+\xi Q_{ext}^{\prime}(X,t).
\end{eqnarray}
Equation (40) contains both slow and fast part and they must be separately equal. Thus
\begin{equation}
m\ddot{\xi}+\gamma_0\dot{\xi}=F_{ext}(X,t).
\end{equation}
For $F_{ext}(X,t)=f(X)\cos(\Omega t)+g(X)\sin(\Omega t)$, we may choose
\begin{eqnarray}
\hskip-1.8cm
\xi(X,t)&=& -\frac{1}{m(\Omega^2+\frac{\gamma_0^2}{m^2})}\Big\lbrack(f(X)+\frac{\gamma_0}{m\Omega}g(X))\cos(\Omega t) \nonumber \\
&&+ (g(X)-\frac{\gamma_0}{m\Omega}f(X))\sin(\Omega t)\Big\rbrack.
\end{eqnarray}
Now, putting Eq. (42) in Eq. (40) and averaging over one time-period (which is denoted by subscript $\tau$) of the external field, we obtain
\begin{eqnarray}
\hskip-1.8cm
m\ddot{X}(t)+\gamma_0\dot{X}=-\frac{\partial V_0(X)}{\partial X}+\langle \xi F_{ext}^{\prime}(X,t)\rangle_{\tau}+\langle Q_{ext}(X,t)\rangle_{\tau}+\langle \xi Q_{ext}^{\prime}(X,t)\rangle_{\tau}.
\end{eqnarray}
Now,
\begin{eqnarray}
\hskip-1.8cm
\langle \xi F_{ext}^{\prime}(X,t)\rangle_{\tau}&=&-\frac{1}{4m(\Omega^2+\frac{\gamma_0^2}{m^2})}\Big\lbrack\frac{\partial}{\partial X}(f^2(X)+g^2(X))\nonumber \\
&&+\frac{2\gamma_0}{m\Omega}(f^{\prime}(X)g(X)-g^{\prime}(X)f(X))\Big\rbrack.
\end{eqnarray}
\begin{eqnarray}
\hskip -1.8cm
\langle Q_{ext}(X,t)\rangle_{\tau}=-\frac{1}{2T}\int_0^Tdt\lbrace f^{\prime\prime\prime}(X)\cos(\Omega t)+g^{\prime\prime\prime}(X)\sin(\Omega t)\rbrace\Big\lbrack \langle \delta \hat{x}^2\rangle_0\Big\rbrack.
\end{eqnarray}
Since the term in square bracket is independent of time (see Eq. (35)), we obtain $\langle Q_{ext}(X,t)\rangle_{\tau}=0$. In the same manner one can show that in the limit of large $t>\gamma_0^{-1}$
\begin{eqnarray}
\langle \xi Q_{ext}^{\prime}(X,t)\rangle_{\tau}&=&-\frac{1}{2m(\Omega^2+\frac{\gamma_0^2}{m^2})}\Big\lbrack\langle \delta \hat{x}^2\rangle\Big\rbrack_{0}\Big\lbrace(f(X)f^{\prime\prime\prime}(X)+g(X)g^{\prime\prime\prime}(X))\nonumber \\
&+&\frac{\gamma_0}{m\Omega}\Big(f^{\prime\prime\prime}(X)g(X)-g^{\prime\prime\prime}(X)f(X))\Big\rbrace,
\end{eqnarray}
where we have replaced $\langle \delta \hat{x}^2(t)\rangle_0$ by its equilibrium value $\langle \delta \hat{x}^2\rangle_{0}$. Finally, one can write down the effective equation of motion for the slow variable as follows
\begin{equation}
m\ddot{X}+\gamma_0\dot{X}=-\frac{\partial U_{eff}}{\partial X},
\end{equation}
where 
\begin{equation}
 U_{eff} = V_0+V_{cl}+V_{QD},
\end{equation}
\begin{eqnarray}
\hspace{-1.8cm}
V_{cl}&=&\frac{1}{4m(\Omega^2+\frac{\gamma_0^2}{m^2})}\Big\lbrack(f^2(X)+g^2(X))
+\frac{2\gamma_0}{m\Omega}\int_0^{X}dy(g(y)f^{\prime}(y)-f(y)g^{\prime}(y))\Big\rbrack,\nonumber \\
\hspace{-1.8cm}
&=& V_{KLL}+V_{CD},
\end{eqnarray}
with $V_{KLL}=\frac{(f^2(X)+g^2(X))}{4m(\Omega^2+\frac{\gamma_0^2}{m^2})}$ and $V_{CD}=\frac{\gamma_0}{2m^2\Omega(\Omega^2+\frac{\gamma_0^2}{m^2})}\int_0^{X}dy(g(y)f^{\prime}(y)-f(y)g^{\prime}(y))$.
Finally, the quantum contribution is given by :
\begin{eqnarray}
\hspace{-0.8cm}
V_{QD}&=& \frac{1}{4m(\Omega^2+\frac{\gamma_0^2}{m^2})}\Big\lbrack\langle\delta\hat{x}^2\rangle\Big\rbrack_{0}\Big\lbrace\int_0^X dy(f(y)f^{\prime\prime\prime}(y)+g(y)g^{\prime\prime\prime}(y))\nonumber \\
\hspace{-0.8cm}
&&+\frac{\gamma_0}{m\Omega}\int_0^X dy(g(y)f^{\prime\prime\prime}(y)-f(y)g^{\prime\prime\prime}(y))\Big\rbrace.
\end{eqnarray}
Equation (50) embodies the leading order quantum correction. In the classical limit, $V_{QD}$ vanishes and the effective potential reduces to $V_0+V_{KLL}+V_{CD}$, which is the result of \cite{barma,malay}. We also note that in the absence of dissipation  $\gamma_0\rightarrow 0$, both $V_{CD}$ and $V_{QD}$ vanish and the effective potential reduces to $V_0+V_{KLL}$, consistent with the results of \cite{cook,rahav}. Finally, in the limit of very high dissipation $\gamma_0\rightarrow \infty$, the quantum correction is negligible as is evident from Eq. (35). Thus, we conclude that the effects of quantum dissipative contribution are largest for intermediate values of damping.\\
\indent
To illustrate the evolution of the form of $V_{eff}$ as the damping is varied, we consider a system which is perturbed by an oscillatory potential with a Gaussian profile i.e. $V_{ext}(\hat{x},t)=\Gamma \exp(- \hat{x}^2/a^2)\cos(\Omega t)$. The effective potential then has the following parts : 
\begin{eqnarray}
\tilde{V}_0 & = &=\frac{ V_0}{E_0}=\tilde{X}^2, \\
\tilde{V}_{cl} & = &=\frac{V_{cl}}{E_0}=\frac{A}{1+\frac{\gamma_0^2}{m^2\Omega^2}}\tilde{X}^2\exp(-2\tilde{X}^2),\\
\tilde{V}_{QD} & = &=\frac{V_{QD}}{E_0}=\frac{A}{\sqrt{\pi}(1+\frac{\gamma_0^2}{m^2\Omega^2})}\lbrack\langle\tilde{\delta}\hat{x}^2\rangle\rbrack_{0}\Big\lbrack 6\Big(\Gamma(2,2\tilde{X}^2)-\Gamma(2)\Big)\nonumber \\
&&+3(1-e^{-2\tilde{X}^2})-\Big(\Gamma(3,2\tilde{X}^2)-\Gamma(3)\Big)\Big\rbrack,
\end{eqnarray}
where $\Gamma(b,y)$ is the incomplete gamma function and $E_0=\frac{1}{2}m\omega_0^2a^2$. 
\begin{figure}[h]
\begin{center}
{\rotatebox{270}{\resizebox{6cm}{12cm}{\includegraphics{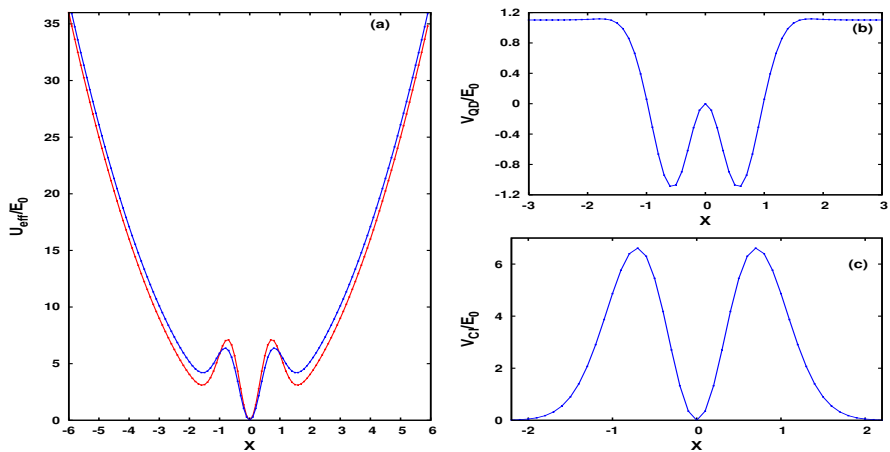}}}}
\caption{(color online) Plot of the effective time independent potential versus $X$ for very weak damping with external driving  potential $V_{ext}(x,t)=\Gamma  \exp(- x^2/a^2)\cos(\Omega t)$ (a) without (red filled circle; $\tilde{V}_0+\tilde{V}_{cl}$) and with (blue filled square; $\tilde{V}_0+\tilde{V}_{cl}+\tilde{V}_{QD}$) quantum dissipative part, (b) only quantum dissipative part ($\tilde{V}_{QD}$) and (c) only classical part ($\tilde{V}_{cl}$). To plot this figure we used $\gamma_0/m\omega_0=0.08$, $\Omega/\omega_0=150$, $A=17.85$, $a/\sqrt{\frac{\hbar}{m\omega_0}}=0.71$,  and $k_BT/\hbar\omega_0=0.1$. We have set $\hbar=m=k_B=1$.}
\end{center}
\end{figure}
The above equations are expressed in terms of dimensionless variables $\tilde{X}=X/a$, $A=\frac{\Gamma^2}{16mE_0\Omega^2a^2}$, and $\lbrack\langle\tilde{\delta}\hat{x}^2\rangle\rbrack_{0}= \lbrack\langle{\delta}\hat{x}^2\rangle\rbrack_{0}/a^2$. 
\begin{figure}[h]
\begin{center}
{\rotatebox{270}{\resizebox{6cm}{12cm}{\includegraphics{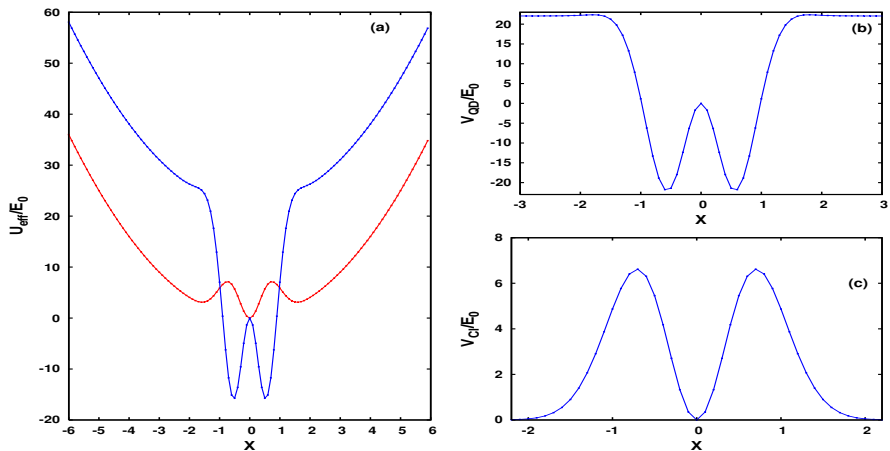}}}}
\caption{(color online) Plot of the effective time independent potential versus $X$ for moderate damping values with external driving and parameters as in Figure 1, except $\gamma_0/m\omega_0=1.0$.}
\end{center}
\end{figure}
\begin{figure}[h]
\begin{center}
{\rotatebox{270}{\resizebox{6cm}{12cm}{\includegraphics{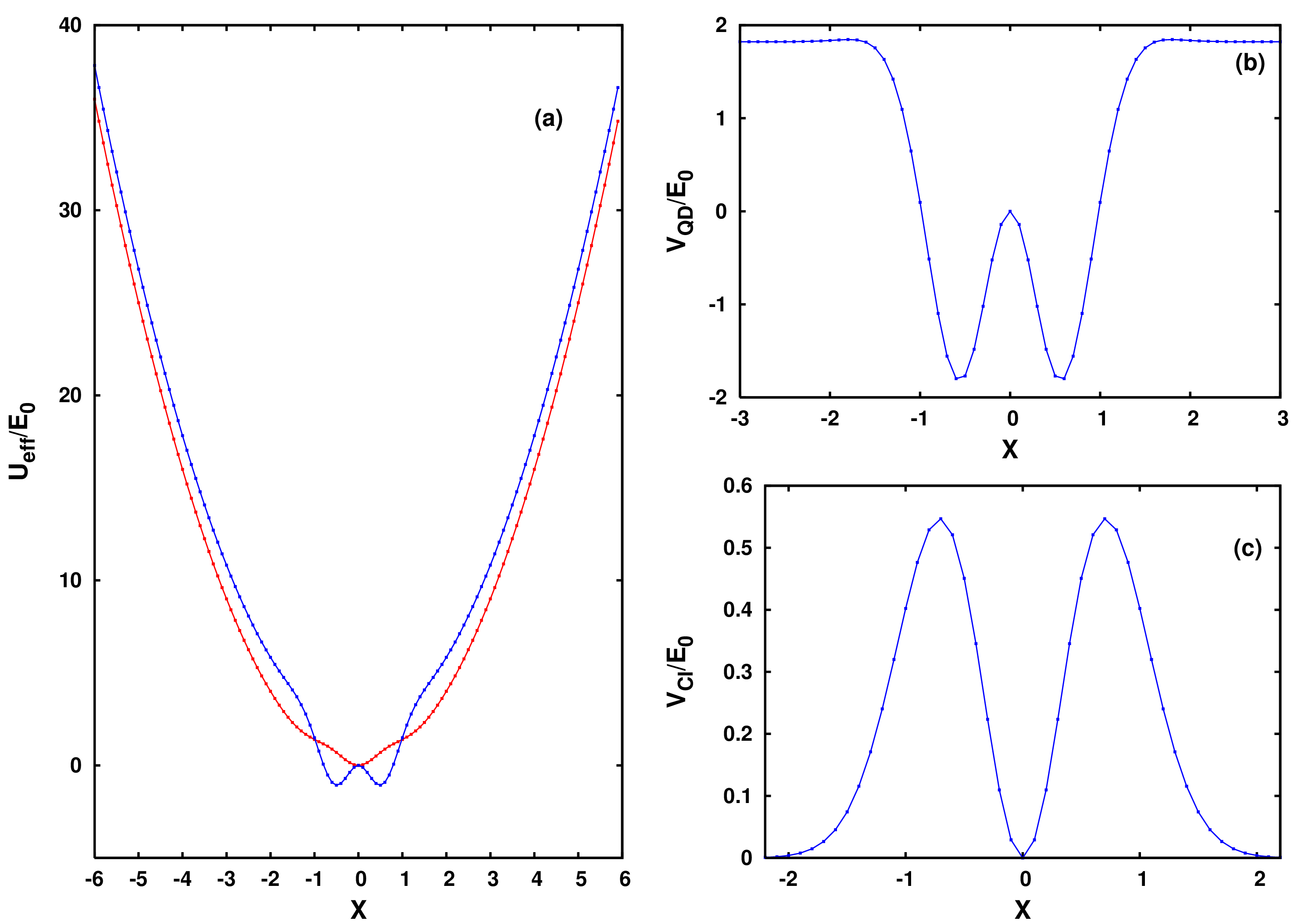}}}}
\caption{(color online) Plot of the effective time independent potential versus $X$ for high damping values with external driving and parameters as in Figure 1, except  $\gamma_0/m\omega_0=10.0$.}
\end{center}
\end{figure}
Figures (1), (2), and (3) depicts the effective potential profile in the low, intermediate and high damping cases respectively. As remarked in the previous paragraph, we see that the effects of $V_{QD}$ are most pronounced at intermediate values of the damping. The presence of higher order derivatives of space dependent terms of forcing in the quantum dissipative part, $V_{QD}$, brings in some interesting changes in the qualitative form of the effective potential. It is evident from figure (1) that in the absence of dissipation the effective potential has two metastable minima and one stable minimum . But, on adding $V_{QD}$ two metastable states become stable minima and the stable minimum becomes a maximum.\\
\section{Conclusion}
In this paper, we have analyzed the properties of a quantum particle in a harmonic potential, coupled to a dissipative Ohmic heat bath, in the presence of a space dependent rapidly oscillating field. Starting from the usual heat-bath reservoir model, we obtained a $c$-number GQLE  from the operator equivalent by using a coherent state representation of bath oscillators \cite{banerjee1,banerjee2}. This enabled us to apply the Kapitza idea of separating into a slow part and a fast part, which consists of the rapid oscillations  around the slow motion. The resulting equation for the slow motion is treated perturbatively. To leading order, we find that the rapid forcing leads to an effective static potential which has a quantum contribution, $V_{QD}$, which adds to the classical result. This quantum contribution vanishes for both very small and very large values of the damping, but can have an appreciable and interesting effects for intermediate values. For instance, the number of stable minima in the effective potential can change, as illustrated by the case of a Gaussian oscillating force, for moderate values of the damping.\\

\end{document}